\documentclass[aps,pre,amssymb,amsmath,twocolumn,showpacs]{revtex4-1}

\usepackage{graphicx}
\usepackage{dcolumn}
\usepackage{bm,dsfont}
\usepackage{epsfig}
\usepackage{epstopdf}
\usepackage[T1]{fontenc}
\usepackage{lmodern}

\usepackage{hyperref}

\usepackage{color}
\definecolor{dblue}{rgb}{0,0,0} 
\long\def\blue#1{{\color{dblue}{#1}\color{black}}}

\begin{document}
\title{Pair approximation for the $q$-voter model with independence on complex networks}
\author{Arkadiusz J\k{e}drzejewski} 
\affiliation{Department of Theoretical Physics, Wroc\l{}aw University of Science and Technology, Wroc\l{}aw, Poland}
\date{\today}
\email{arkadiusz.jedrzejewski@pwr.edu.pl}
\begin{abstract}
We investigate the $q$-voter model with stochastic noise arising from independence on complex networks. Using the pair approximation, we provide a comprehensive, mathematical description of its behavior and derive a formula for the critical point. The
analytical results are validated by carrying out Monte Carlo experiments.
The pair approximation prediction exhibits substantial agreement with simulations, especially for networks with weak clustering and large average degree. Nonetheless, for the average degree close to $q$, some discrepancies originate.
It is the first time we are aware of that the presented approach has been applied to the nonlinear voter dynamics with noise.
Up till now, the analytical results have been obtained only for a complete graph.
We show that in the limiting case the prediction of pair approximation coincides with the known solution on a fully connected network.\\\\
Post-print of A. J\k{e}drzejewski, \href{https://doi.org/10.1103/PhysRevE.95.012307}{Phys. Rev. E \textbf{95}, 012307 (2017)}.\\
Copyright (2017) by the American Physical Society.
\end{abstract}
\maketitle
\section{Introduction}
Investigating diffusion processes is a subject of broad and current interest \cite{pastor2015epidemic}. 
Especially important is to comprehend their dynamical behavior and understand which factors \blue{are significant for the evolution of a system.}
This information may turn out to be critical for
epidemiologists in preventing contagious diseases and establishing health policies. However,
applications of dynamical models reach far beyond that. In computer science, they
describe propagation of data packets or malicious software and help to design routing
protocols and more robust networks \cite{callaway2000network}. Information search and retrieval are relevant to the
dynamics of spreading, as well. In engineering, similar models are used and analyzed in
order to plan efficient power grids and regulate energy policies as well as study conductivity
and flow through porous materials \cite{hunt2014percolation}. Sociologists in the same terms describe phenomena like opinion
dynamics, emergence of consensus, and diffusion of innovations \cite{squazzoni2010impact}.

In recent years, especially considerable attention has been drawn to the $q$-voter model \cite{javarone2015conformism,jedrzejewski2016mapping,siedlecki2016interplay,mellor2016characterization,mobilia2015nonlinear,timpanaro2014exit,timpanaro2015analytical,chmiel2015phase,sznajd2014nonlinear}, which was originally proposed by
Castellano \textit{et al.} \cite{castellano2009nonlinear} and later altered by Nyczka \textit{et al.} \cite{nyczka2012}. Its modified
version represents opinion dynamics under two types of social response: conformity and independence \cite{nail2016diamond}. Although widely exploited in sociophysics, it also finds application
in computational economics as an underlying mechanism of consumers decision making
process, sometimes with slight modifications. Recently, it was adopted in the analysis
of designing marketing strategies and pricing innovative commodities \cite{kowalska2016linking,maciejowska2016impact}. Byrka \textit{et
al.} \cite{byrka2016difficulty}, on the other hand, proposed similar agent-based model with $q$-voter dynamics in
order to explain a behavioral gap between people's intentions and practices.

Although researchers associate the $q$-voter model with opinion dynamics, it is frequently allocated on simple topologies like complete graphs or regular lattices \cite{mellor2016characterization,mobilia2015nonlinear,timpanaro2014exit,timpanaro2015analytical} despite the fact that social networks are regarded as complex structures \cite{boccaletti2006complex}. 
And even if some studies consider irregularities and randomness in modeling pairwise relations between agents, they focus primarily on simulation results, and only minority proposes analytical approach, as well \cite{moretti2013mean,vazquez2008analytical,sood2008voter}.
Thus, in this paper, we carry out a mathematical analysis into the
problem, and we clearly establish a connection between $q$-voter dynamics and complex networks.
The system is studied by using pair approximation, which was previously employed only to the linear voter model \cite{vazquez2008analytical} and has not been applied before to the nonlinear one.
Moreover, the paper refers to the $q$-voter dynamics with stochastic noise, sociologically interpreted as independence, for which analytical results are known only in the case of a complete graph \cite{nyczka2012}.
The obtained approximation is validated by carrying out Monte Carlo simulations on several structures like random regular (RR), Erd\H{o}s-R\'{e}nyi (ER), Watts-Strogatz (WS), and scale-free (SF) networks, including the Barab\'{a}si-Albert (BA) model.
We also show that in the limiting case our prediction coincides with the solution for a complete graph.

\section{model description}
We consider an arbitrary network of the size $N$. Each vertex is associated with one autonomous agent characterized by a binary, spin like variable $s_i=\pm1$.
In every elementary time step, we pick at random an agent.
\blue{In the original voter dynamics \cite{clifford1973model}, the agent embraces the same opinion of its randomly selected neighbor.
This simple rule leads to a linear form of the flipping probability $f(x)=x$, i.e., the probability that considered spin flips surrounded by a fraction $x$ of neighbors in the opposite state.
Non-linearity in the $q$-voter model, on the other hand, is introduced by engaging a group of $q$ individuals, which influences the current state of an agent.
As a result, the flipping probability does not have a linear form as in the ordinary voter model \cite{castellano2009nonlinear}.}
The group is called $q$-panel, and its potential members are determined by the network topology and selected at random.
With probability $p$, the chosen agent acts independently and adopts the opposite opinion or preserves the old one with equal chances. Otherwise, with probability $1-p$, it behaves like a conformist and embraces the viewpoint of $q$-panel, but only if the group is unanimous, that is to say, all $q$ individuals have the same state. 
Drawing agents occurs without repetition so that the above definition of the model is identical to that in Ref.~\cite{nyczka2012}. Nevertheless, now we extend underlying topologies to more complex structures than complete graphs.
\blue{In simulations, we ensure that all vertices have at least $q$ neighbors by selecting appropriate parameters of investigated networks so that each agent could be exposed to group influence.}

One of the macroscopic quantities, which is easily traceable and can characterize our system, is a concentration or a fraction of up-spins: 
\begin{equation}
    c_{\uparrow}=\frac{1}{2N}\sum_{i=1}^N (1+s_i).
    \label{eq:upc}
\end{equation}
It can be interpreted as a probability of finding an agent with a positive opinion, namely $s_i=1$.
Analogously, we can define a concentration of down spins
\begin{equation}
    c_{\downarrow}=\frac{1}{2N}\sum_{i=1}^N (1-s_i).
    \label{eq:downc}
\end{equation}
Since randomly choosing an up-spin and a down-spin are complementary events, we have $c_{\uparrow}+c_{\downarrow}=1$. For a notation simplicity, $c$ without index also refers to $c_{\uparrow}$.
Finally, we are also interested in interfaces between any two connected agents. 
In statistical physics, a bond is said to be active if it links two spins with opposite values \cite{sznajd2010phase}.
Herein we adopt the same idea and define a concentration of active bonds as follows:
\begin{equation}
   b=\frac{1}{2N\langle k\rangle}\sum_{i=1}^N\sum_{j=1}^N (1-s_is_j)\mathds{1}_{(i,j)\in E},
    \label{eq:bonds}
\end{equation}
where $\langle k\rangle$ is the average degree of the network, and $\mathds{1}_{(i,j)\in E}$ is an indicator function specified on a set of edges $E$.

Let us consider an agent placed on a vertex that has $k$ neighbors and $i$ active bonds attached. 
The flipping probability \cite{castellano2009nonlinear}, that is to say, the probability that an agent changes its opinion to the opposite one, is a function of active bonds number $i$ connected to agent's node. 
Since we do not allow repetition of chosen agents in a $q$-panel, this probability takes the following form:
\begin{equation}
    f_k(i,q,p)=(1-p)\frac{\prod_{j=1}^q(i-j+1)}{\prod_{j=1}^q(k-j+1)}+\frac{p}{2}.
    \label{eq:flipProb}
\end{equation}

As we mentioned before, the probability of choosing an agent with a positive opinion from the system $P(s_i=1)$, or for the simplicity $P(\uparrow)$, is equal to up-spins concentration,
\begin{equation}
    P(\uparrow)=c_\uparrow=c.
    \label{eq:cup}
\end{equation}
The same holds for choosing an agent with a negative opinion,
\begin{equation}
    P(\downarrow)=c_\downarrow=1-c.
    \label{eq:cdown}
\end{equation}
The concentration of active bonds, on the other hand, possesses the information about mutual relations between agents.
If we think of choosing an active bond as a consecutive selection of two connected nodes, two events may occur. 
First, we select an up-spin and then a down-spin $(\downarrow\uparrow)$, or the other way around, at the beginning, we encounter a down-spin and after an up-spin $(\uparrow\downarrow)$.
As a consequence, one can write the following equation:
\begin{equation}
    b=P(\uparrow\downarrow)+P(\downarrow\uparrow).
    \label{eq:bprob}
\end{equation}
Moreover, there is the same amount of connections between $\uparrow\downarrow$ and $\downarrow\uparrow$ because they share the same edge.
This implies that $P(\uparrow\downarrow)=P(\downarrow\uparrow)$ and together with Eq.~(\ref{eq:bprob}), we obtain
\begin{equation}
    P(\uparrow\downarrow)=P(\downarrow\uparrow)=\frac{b}{2}.
    \label{eq:andProb}
\end{equation}
Having the above result, calculating conditional probabilities is straightforward. 
We use the following notation $P(\downarrow|\uparrow)$ to denote the probability of selecting a down-spin given we have already picked an up-spin and so on for other spin combinations:
\begin{equation}
\begin{split}
    P(\downarrow|\uparrow)&=\frac{P(\downarrow\uparrow)}{P(\uparrow)}=\frac{b}{2c},\\
    P(\uparrow|\downarrow)&=\frac{P(\uparrow\downarrow)}{P(\downarrow)}=\frac{b}{2(1-c)}.
    \label{eq:conditionalProb}
\end{split}
\end{equation}
In order to simplify further notation we designate \mbox{$P(\downarrow|\uparrow)$} by $\theta_\uparrow$
and \mbox{$P(\uparrow|\downarrow)$} by $\theta_\downarrow$.

\section{Pair approximation}
In an elementary time step, only a single agent may change his opinion.
This event affects concentration of active bonds as well as concentration of up spins.
Let us consider first the fraction of agents with a positive attitude.
There are three possible scenarios. 
The amount of up spins may increase by $1$, decrease by $1$, or remain at the same level.
So a flip corresponds to an elementary change in $c$ by $\Delta_c=1/N$.
Following Ref.~\cite{nyczka2012}, we denote probabilities of these changes by
\begin{equation}
\begin{split}
    \gamma^+ &=P(c\rightarrow c+\Delta_c),\\
    \gamma^- &=P(c\rightarrow c-\Delta_c).\\
\end{split}
\end{equation}
Knowing the above transition probabilities, we are able to write down the rate equation \cite{krapivsky2010kinetic}, which describes the dynamics of up-spins concentration,
\begin{equation}
    c(t+\Delta t)=c(t)+\Delta_c(\gamma^+-\gamma^-).
    \label{eq:cdisc}
\end{equation}
When we deal with large systems $N\gg1$, in particular, when $N\to\infty$, the time interval tends to zero $\Delta t\to 0$.
In such a case Eq.~(\ref{eq:cdisc}) takes continuous form,
\begin{equation}
    \frac{\partial c}{\partial t}=\gamma^+-\gamma^-.
    \label{eq:partialC}
\end{equation}
Usually, the explicit formulas for $\gamma^+$ and $\gamma^-$ are derived based on the mean-field approximation (MFA) as in Refs.~\cite{nyczka2012,mobilia2015nonlinear,castellano2009nonlinear}.
This theory postulates agents homogeneity in order to reduce the complexity of a problem. 
In more detail, it neglects all fluctuations in a system so that a local concentration is equal to the global, average one.
This approach gives prefect results on a complete graph where, indeed, our system is homogeneous and isotropic.
However, on more complicated structures, not taking into account fluctuations gives overestimated results.

Herein, we adopt the improved method called the pair approximation (PA) \cite{vazquez2008analytical,gleeson2013binary}, which also considers dynamical correlations between the nearest neighbors; therefore, it delivers more precise local description of a system than the ordinary mean-field approach.
This additional information is acquired from the active bonds concentration through conditional probabilities derived before.
In PA, we assume that states of a chosen agent's neighbors are independent of each other; thus, 
$P(\uparrow |\downarrow)$ and $P(\downarrow|\uparrow)$ remain the same for all spins in the panel.
This, however, may not be true in general, especially when a network is highly clustered.
Furthermore, we postulate no degree correlations, i.e., we neglect all dependencies between node degree and finding a spin in a given position or a link in a given state.
So no matter what is the value of $k$ for a picked vertex, we use the same $c$ and $b$, which characterize all nodes.

Let us derive transition probabilities relying on PA.
The reasoning for $\gamma^+$ is following.
The concentration of up-spins has a chance to increase only if an agent with negative attitude is picked.
This event happens with probability $P(\downarrow)$.
Then an agent is submitted to either conformity or independence. 
In the former case, a unanimous $q$-panel of up-spins $\uparrow$ must be gathered.
Choosing each of them is $P(\uparrow | \downarrow)\equiv\theta_\downarrow$ probable since we already know the opinion of the considered agent.
In the latter case, a flip occurs with probability $1/2$.
Analogical inference might be applied to $\gamma^-$. As a result we obtain
\begin{equation}
\begin{split}
    \gamma^+ &=(1-c)\left[(1-p)\theta_\downarrow^q+\frac{p}{2}\right],\\
    \gamma^- &=c\left[(1-p)\theta_\uparrow^q+\frac{p}{2}\right].
\end{split}
\label{eq:gammas}
\end{equation}
Nevertheless, we are still not able to solve Eq.~(\ref{eq:partialC}) without knowing the form of $b$ which is necessary for the conditional probabilities $\theta_\downarrow$ and $\theta_\uparrow$.

For the concentration of active bonds, we can conduct similar calculations and also find its rate equation, following the reasoning of Ref.~\cite{vazquez2008analytical}.
This time, a single flip refers to an elementary change in $b$ by $\Delta_b$ dependent on agent's node degree and number of already active edges $i$ attached to this vertex.
Locally, after a flip, links that were inactive become active and oppositely.
Remembering that the total amount of edges in a network equals $N\langle k\rangle/2$, we obtain
\begin{equation}
    \Delta_b=\frac{2}{N\langle k\rangle}(k-2i),
    \label{eq:smalDeltab}
\end{equation}
where $i$ ranges from $0$ to $k$.
Let us explicitly write down the overall likeliness of changing an agent's opinion, which is just a sum of transition probabilities $\gamma^++\gamma^-$ in the following form:
\begin{equation}
    P(c\rightarrow c\pm\Delta_c)=\sum_{j\in \{ \uparrow,\downarrow \}}P(c\rightarrow c\pm\Delta_c|j)P(j),
    \label{eq:part1}
\end{equation}
where $P(c\rightarrow c\pm\Delta_c|j)$ is the conditional probability of a flip given that the chosen spin is in a state \mbox{$j\in \{ \uparrow,\downarrow \}$}; see Eqs.~(\ref{eq:gammas}).
Furthermore, using our assumption that the node degree is independent of an agent's opinion and applying the law of total probability, the following equation is obtained:
\begin{equation}
    P(c\rightarrow c\pm\Delta_c|j)=\sum_k P(c\rightarrow c\pm\Delta_b|k,j)P(k).
    \label{eq:part2}
\end{equation}
Among all $k$ edges attached to the picked vertex, $i$ of them might be active where $i\in\{0,1,...,k\}$.
The probability of selecting one of them is $P(-j|j)\equiv\theta_j$, and since we treat these events as independent, picking $i$ of active bonds is binomially distributed $B(k,\theta_j)$. 
It results in
\begin{align}
    P(c\rightarrow c\pm\Delta_c|k,j)=&\sum_{i=0}^kP(c\rightarrow c\pm\Delta_c|k,j,i)\nonumber\\
    &\times\binom{k}{i}\theta_j^i(1-\theta_j)^{k-i},
    \label{eq:part3}
\end{align}
where under $-j$ we understand the opposite state to $j$.
Finally, we obtain the formula in which the conditional probability of a flip $P(c\rightarrow c\pm\Delta_c|k,j,i)$ is used given the number of active bonds, the chosen agent's opinion, and the node degree.
But this, in fact, is the flipping probability $f_k(i,q,p)$ from Eq.~(\ref{eq:flipProb}).
Now, having this partition we know the impact of elementary shifts in $b$ for different events.
Each flip that is $f_k(i,q,p)$ probable causes a change in the overall active bonds concentration $\Delta b$ by $\Delta_b$; thus, combining Eqs.~(\ref{eq:part1})-(\ref{eq:part3}) we get
\begin{align}
    \Delta b=&\sum_{j\in\{\uparrow,\downarrow\}}c_j\sum_k P(k)\sum_{i=0}^{k}\binom{k}{i}\theta_j^i(1-\theta_j)^{k-i}\nonumber\\
    &\times f_k(i,q,p)\Delta_b.
\end{align}
Connecting this result with Eq.~(\ref{eq:smalDeltab}) and the fact that $\Delta t=1/N$, the following expression is obtained:
\begin{align}
    \frac{\Delta b}{\Delta t}=&\frac{2}{\langle k\rangle}\sum_{j\in\{\uparrow,\downarrow\}}c_j\sum_k P(k)\sum_{i=0}^{k}\binom{k}{i}\theta_j^i(1-\theta_j)^{k-i}\nonumber\\
    &\times f_k(i,q,p)(k-2i).
\end{align}
Of course, once again we can take a limit $N\rightarrow\infty$ and attain the differential form.
Altogether, we obtain a system of two differential equations that can be solved simultaneously.
The first for up-spins and the second one for active bonds concentration:

\begin{align}
    \frac{\partial c}{\partial t}=&\gamma^+-\gamma^-, \nonumber\\
    \frac{\partial b}{\partial t}=&\frac{2}{\langle k\rangle}\sum_{j\in\{\uparrow,\downarrow\}}c_j \nonumber\\
    &\times\sum_k P(k)\sum_{i=0}^{k}{{k}\choose{i}}\theta_j^i(1-\theta_j)^{k-i}f_k(i,q,p)(k-2i).
    \label{eq:system}
\end{align}

The above result is general and can be applied to other dynamical models from the same class.
Then the forms of transition and flipping probabilities will alter, but Eq.~(\ref{eq:system}) will remain unchanged.

Let us simplify the formula for the active bonds concentration in the above system of equations for the $q$-voter model.
First, notice that the flipping probability can be presented in the equivalent form:
\begin{equation}
    f_k(i,q,p)=(1-p)\frac{i!(k-q)!}{k!(i-q)!}+\frac{p}{2}.
    \label{eq:flipProb2}
\end{equation}
Now, we consider only the interior sum over active bonds~$i$ in Eq.~(\ref{eq:system}), and we decompose it into two summations:
\begin{align} 
    k&\sum_{i=0}^{k}\binom{k}{i}\theta_j^i(1-\theta_j)^{k-i}f_k(i,q,p)\nonumber\\
    -2&\sum_{i=0}^{k}\binom{k}{i}\theta_j^i(1-\theta_j)^{k-i}f_k(i,q,p)i.
    \label{eq:innerSum}
\end{align}
After substitution Eq.~(\ref{eq:flipProb2}) for $f_k(i,q,p)$, we have four terms to be calculated
\begin{align}
    k(1-p)&\sum_{i=q}^{k}\binom{k}{i}\theta_j^i(1-\theta_j)^{k-i}\frac{i!(k-q)!}{k!(i-q)!}\nonumber\\
    +k\frac{p}{2}&\sum_{i=0}^{k}\binom{k}{i}\theta_j^i(1-\theta_j)^{k-i}\nonumber\\
    -2(1-p)&\sum_{i=q}^{k}\binom{k}{i}\theta_j^i(1-\theta_j)^{k-i}\frac{i!(k-q)!}{k!(i-q)!}i\nonumber\\
    -p&\sum_{i=0}^{k}\binom{k}{i}\theta_j^i(1-\theta_j)^{k-i}i.
    \label{eq:term1}
\end{align}
\begin{figure}[t!]
	\centerline{\epsfig{file=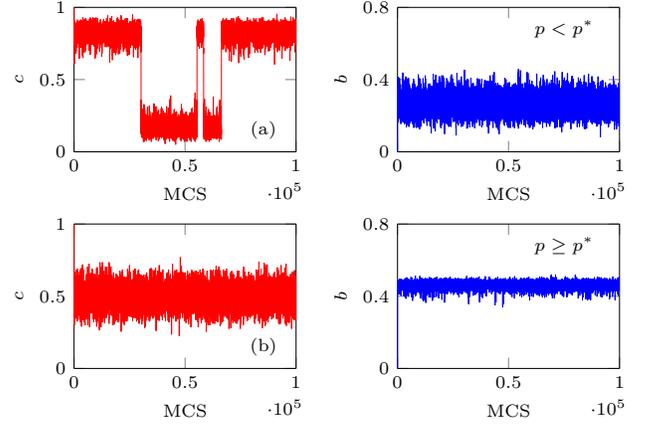,width=1.0\columnwidth}}
	\caption{\label{fig:timeEvolution2q}(Color online) Typical, sample trajectories of the up spins concentrations and the corresponding active bonds concentrations when $q\leq 5$. For the presented realizations $q=2$, and the network topology is a random $14$-regular graph of the size $N=200$. 
		The level of independence: (a) $p=0.2$ and (b) $p=0.4$. Spontaneous transitions between stationary states are noticeable below the critical point $p<p^*$ and only for $c$.}
\end{figure}
Remembering that the active bonds number is a random variable $\sim B(k,\theta_j)$, we see that the last sum is its first moment $k\theta_j$.
The second summation, on the other hand, is equal to $1$ since it is the total probability for all hypothetical outcomes $i$.
The first and the third term we have to compute explicitly; note that here we start addition from $q$, not $0$. That is because for $i<q$ the front term in Eq.~(\ref{eq:flipProb}) vanishes.
The first term calculations are as follows:
\begin{align}
    &\sum_{i=q}^{k}\binom{k}{i}\theta^i_j(1-\theta_j)^{k-i}\frac{i!(k-q)!}{k!(i-q)!}\nonumber\\
    =&\sum_{i=q}^{k}\frac{(k-q)!}{(k-i)!(i-q)!}\theta^i_j(1-\theta_j)^{k-i}\nonumber\\
    =&\sum_{i=q}^{k}\binom{k-q}{i-q}\theta^i_j(1-\theta_j)^{k-i}
    =\theta^{q}_j.
    \label{eq:term2}
\end{align}
The third term calculations are as follows:
\begin{align}
    &\sum_{i=q}^{k}\binom{k}{i}\theta^i_j(1-\theta_j)^{k-i}\frac{i!(k-q)!}{k!(i-q)!}i\nonumber\\
    =&\sum_{i=q}^{k}\frac{(k-q)!}{(k-i)!(i-q)!}\theta^i_j(1-\theta_j)^{k-i}i\nonumber\\
    =&\sum_{i=q}^{k}\binom{k-q}{i-q}\theta^i_j(1-\theta_j)^{k-i}i\nonumber\\
    =&\theta^{q}_j\left.[(k-q)\theta_j+q\right].
    \label{eq:term3}
\end{align}
After combining together Eqs.~(\ref{eq:term1})-(\ref{eq:term3}) and substituting them back into Eq.~(\ref{eq:system}), we can perform the summation over $k$.
In the end, we attain a system of differential equations for the $q$-voter model Eq.~(\ref{eq:qVotersystem}). Although we are not able to solve it analytically, we can do it numerically, for example, using the fourth-order Runge-Kutta scheme:
\begin{align}
    \frac{\partial c}{\partial t}=&(1-p)\left(c_\downarrow\theta_\downarrow^q-c_\uparrow\theta_\uparrow^q\right)+\frac{p}{2}(c_\downarrow-c_\uparrow), \nonumber\\
    \frac{\partial b}{\partial t}=&\frac{2}{\langle k\rangle}\sum_{j\in\{\uparrow,\downarrow\}}c_j\bigg\{(1-p)\theta^{q}_j\left[\langle k\rangle-2q-2\big(\langle k\rangle-q\big)\theta_j\right]\nonumber\\
     &+p\frac{\langle k\rangle}{2}\left(1-2\theta_j\right)\bigg\}.
    \label{eq:qVotersystem}
\end{align}

\section{Time evolution}

\begin{figure}[t!]
    \centerline{\epsfig{file=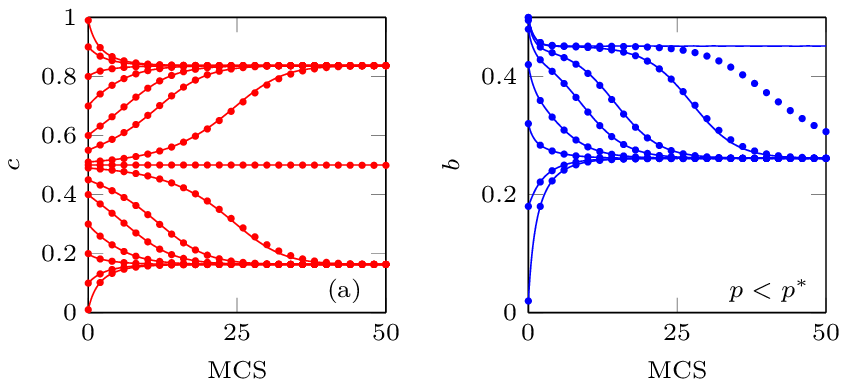,width=1.0\columnwidth}}
    \centerline{\epsfig{file=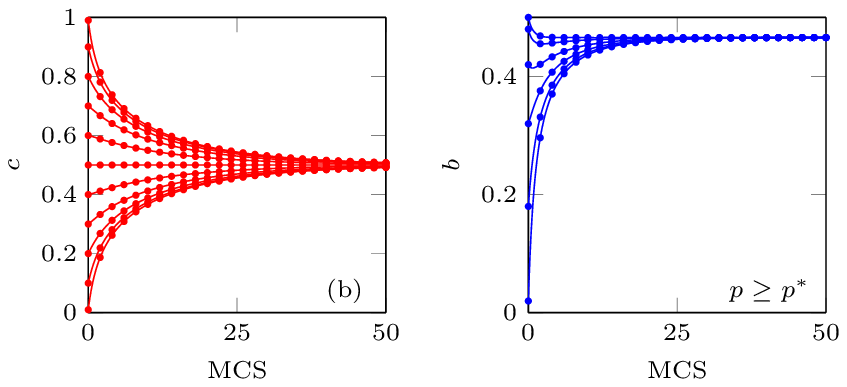,width=1.0\columnwidth}}
    \caption{
    \label{fig:timeEvolution2qBig}(Color online) Average trajectories for a system of the size $N=2\cdot10^5$ on a random $14$-regular graph. In both cases $q=2$. The value of independence: (a) $p=0.2$ and (b) $p=0.35$. Dots represent outcome of the Monte Carlo simulation; solid lines refer to the numerical solution of Eq.~(\ref{eq:qVotersystem}).}
\end{figure}
\begin{figure}[!t]
    \centerline{\epsfig{file=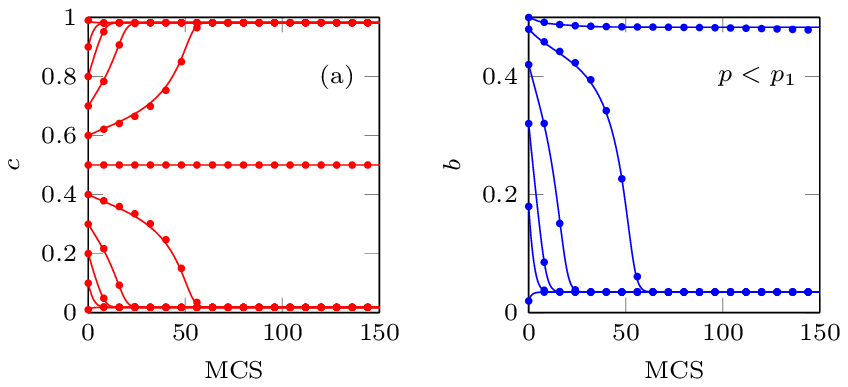,width=1.0\columnwidth}}
    \centerline{\epsfig{file=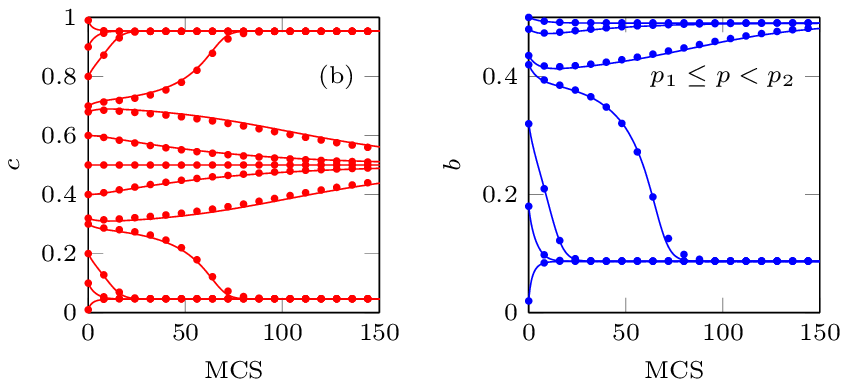,width=1.0\columnwidth}}
    \centerline{\epsfig{file=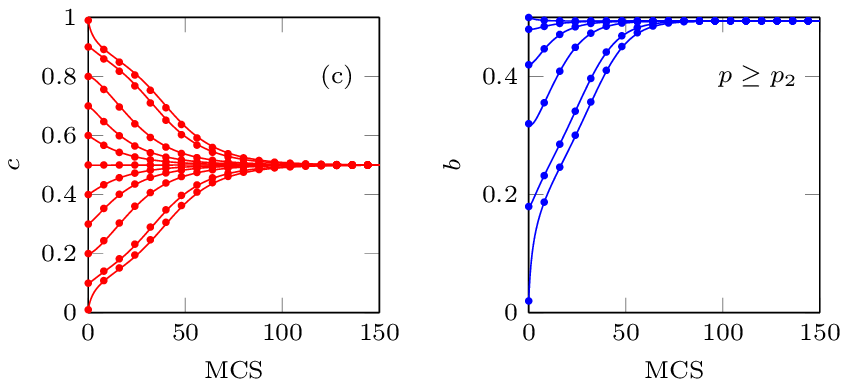,width=1.0\columnwidth}}
    \caption{\label{fig:timeEvolution8qBig}(Color online) Average trajectories for a system placed on BA networks of the size $N=5\cdot10^5$ with $\langle k\rangle=40$ and $q=8$. The level of independence: (a) $p=0.03$, (b) $p=0.06$, and (c) $p=0.1$. Dots represent outcome of the Monte Carlo simulation; solid lines refer to the numerical solution of Eq.~(\ref{eq:qVotersystem}).}
\end{figure}

Figure \ref{fig:timeEvolution2q} presents sample trajectories of the up spins and the active bonds concentrations for different model parameters with the initial condition $c_0=1$.
As seen, the system quickly reaches a steady point and fluctuates around constant values of $c$ and $b$.
However, under certain parameters, there are more stationary states, and occasionally, deviations are strong enough to enforce spontaneous transitions between them.
Along with growing system size, these fluctuations fade, and trajectories seem more like those described by Eq.~(\ref{eq:qVotersystem}).
Eventually, for infinitely large networks, all disturbances disappear.
That is why Figs.~\ref{fig:timeEvolution2qBig} and \ref{fig:timeEvolution8qBig} show the time evolution of $q$-voter model with $N>10^5$ agents, this time also with different initial conditions $c_0$.
Although simulated paths were averaged over 100 realizations, all of them were very alike.
As seen, predicted trajectories almost exactly fit those from Monte Carlo experiment. 
Similarly as for smaller structures, the system reaches a steady point, but now because of its size, it has faint chances of escaping to another steady state.

The qualitative behavior of a system on studied weakly clustered complex networks is similar as on a complete graph and depends on model parameters.
We can distinguish two final, stationary phases -- an ordered one and a disordered one.
They can coexist and create stable and unstable states, it all depends on values of $q$ and $p$.
For $q\leq 5$, there are two stable, ordered phases below the critical level of independence $p^*$.
Above this point, we have only one stable, disordered phase.
It is clearly seen in Fig.~\ref{fig:timeEvolution2q}.
Additionally, when $p<p^*$ we can also distinguish unstable, disordered state present in Fig.~\ref{fig:timeEvolution2qBig}, it corresponds to $c=1/2$. However eventually, because of finite size and fluctuations, the system escapes from it.

When $q\geq 6$, the model behavior is more complicated; see Fig.~\ref{fig:timeEvolution8qBig}. 
Now, we have two characteristic levels of independence $p_1$ and $p_2$.
Above $p_2$, similar to the previous case, we have only one stable, disordered phase, but below, there are three stationary states.
When $p_1\leq p<p_2$, all of them are stable and one of them is disordered.
However, if value of independence is smaller than $p_1$, disordered phase becomes unstable.
Generally, in small systems we can observe spontaneous transitions between stable states.
Unstable ones are noticeable only in larger structures for which fluctuations are smaller.
From all graphs presented herein, it is evident that stationary values of the up spins and the active bonds concentrations are coupled.
In the next section, we will derive analytically this dependency.
\begin{figure}[!t]
    \centerline{\epsfig{file=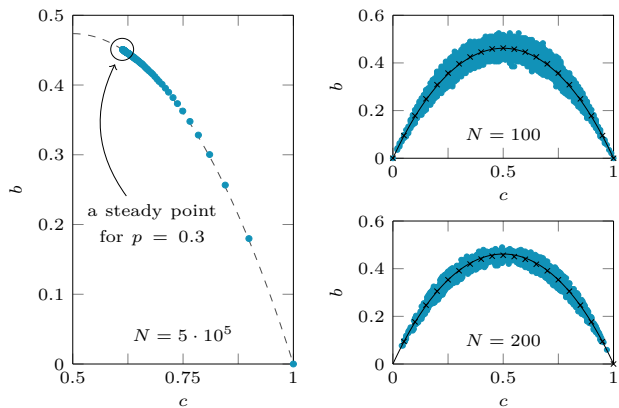,width=1.0\columnwidth,height=5.5cm}}
    \caption{\label{fig:bcdependency}(Color online) The relation between the active bonds and the up-spins concentrations for $q=2$ and different networks: (a) ER model with $\langle k \rangle=20$ and $p=0.3$, (b) and (c) random $14$-regular graphs of different sizes and $p=0.2$. Lines refer to the theoretical result Eq.~(\ref{eq:bdepc}), dots correspond to simulated trajectories starting from the point $(c,b)=(1,0)$. Crosses in the right panels stand for the average value of $b$ for a given $c$.
    Paths consist of 150 MCS for ER and $2\cdot 10^6$ MCS for RR. Fluctuations diminish with the growing system size. }
\end{figure}
\begin{figure}[!b]
	\centerline{\epsfig{file=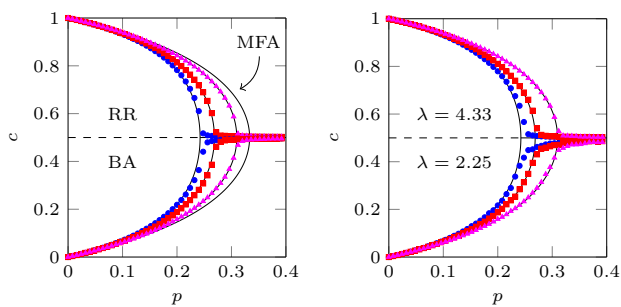,width=1.0\columnwidth}}
	\caption{\label{fig:contPhase}(Color online) Phase diagrams for (a) random regular and Barab\'{a}si-Albert models \blue{and (b) scale-free networks with different exponents. Note that upper and bottom diagrams correspond to different structures, all of them consist of $N=10^5$ vertices.} The group of influence comprises $q=2$ agents. Lines indicate analytical prediction of PA and MFA, solid ones correspond to stable states, and dashed ones refer to unstable points. Markers represent outcomes of Monte Carlo simulations: $\bullet$ $\langle k\rangle=6$, $\scriptstyle\blacksquare$ $\langle k\rangle=8$, $\blacktriangle$ $\langle k\rangle=20$.}
\end{figure}
\section{Stationary states}
In a stationary state, the up-spins as well as active bonds concentration does not change in time, that is to say,
\begin{equation}
    \frac{\partial c}{\partial t}=0\hphantom{00}\wedge\hphantom{00}\frac{\partial b}{\partial t}=0.
    \label{eq:qVotersystemStat}
\end{equation}
Above conditions together with Eq.~(\ref{eq:qVotersystem}) allow us to reveal a connection between $c$ and $b$. 
From the second condition we get
\begin{align}
    0=&\sum_{j\in\{\uparrow,\downarrow\}}c_j(1-p)\theta^{q}_j\left[\langle k\rangle-2q-2\big(\langle k\rangle-q\big)\theta_j\right]\nonumber\\
    &+\sum_{j\in\{\uparrow,\downarrow\}}c_jp\frac{\langle k\rangle}{2}\left(1-2\theta_j\right).
    \label{eq:previousEq}
\end{align}

\begin{figure}[!b]
	\centerline{\epsfig{file=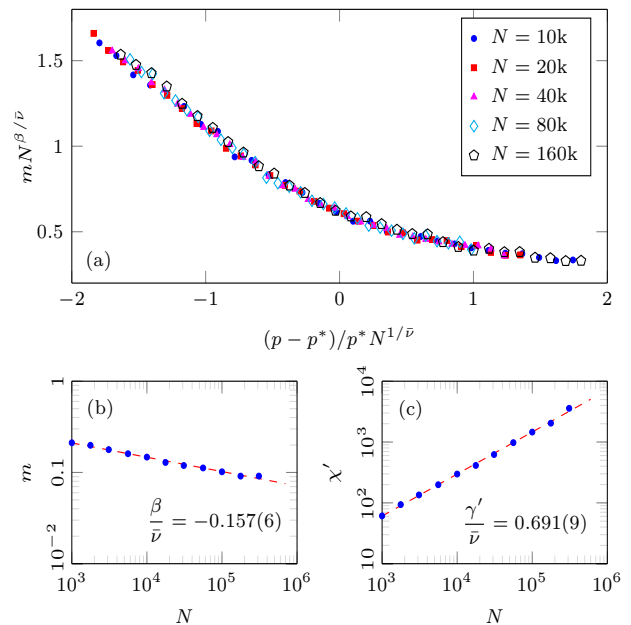,width=1.0\columnwidth,height=8.3cm}}
	\caption{\label{fig:criticalBeh}\blue{(Color online) Critical behavior of the model on scale-free networks with average degree $\langle k\rangle=10$ and tail exponent $\lambda=2.43$. The group of influence comprises $q=2$ agents. (a) A finite size scaling obtained for $\beta/\bar{\nu}=0.157$, $1/\bar{\nu}=0.314$, and $p^*=0.2855$. Critical decays of (b) the magnetization $m$ and (c) the fluctuations $\chi'$ against the system size $N$. Markers refer to Monte Carlo simulations averaged over $1000$ network realizations. Dashed lines, with corresponding slopes $\beta/\bar{\nu}=-0.157(6)$ and $\gamma'/\bar{\nu}=0.691(9)$, come from the linear regression.}}
\end{figure}
Now, from the first condition, let us derive the formula for $p$:
\begin{equation} 
    p=\frac{c_\downarrow\theta_\downarrow^q-c_\uparrow\theta_\uparrow^q}{c_\downarrow\theta_\downarrow^q-c_\uparrow\theta_\uparrow^q-(c_\downarrow-c_\uparrow)/2},
    \label{eq:pEq}
\end{equation}

and insert it into Eq.~(\ref{eq:previousEq}), simultaneously multiplying both sides by the denominator of Eq.~(\ref{eq:pEq}),
\begin{align}
    0=&\sum_{j\in\{\uparrow,\downarrow\}}c_j(c_\uparrow-c_\downarrow)\theta^{q}_j\left[\frac{\langle k\rangle}{2}-q-\big(\langle k\rangle-q\big)\theta_j\right]\nonumber\\
    &+\sum_{j\in\{\uparrow,\downarrow\}}c_j(c_\downarrow\theta_\downarrow^q-c_\uparrow\theta_\uparrow^q)\frac{\langle k\rangle}{2}\left(1-2\theta_j\right).
\end{align}
After preforming summation over $j$, we can get rid of conditional probabilities $\theta_\uparrow$ and $\theta_\downarrow$ as well as the down-spins concentration $c_\downarrow$ by making use of Eqs.~(\ref{eq:cup}), (\ref{eq:cdown}), and (\ref{eq:conditionalProb}).
Finally, after several trivial transformations, we obtain
\blue{
\begin{widetext}
\begin{equation}
    b_{\text{st}}=2\frac{c_{\text{st}}(1-c_{\text{st}})\big[(1-c_{\text{st}})^q-c_{\text{st}}^q\big]-\cfrac{q}{\langle k\rangle}(1-2c_{\text{st}})\big[c_{\text{st}}(1-c_{\text{st}})^q+(1-c_{\text{st}})c_{\text{st}}^q\big]}{\displaystyle(1-c_{\text{st}})^q-c_{\text{st}}^q-\frac{q}{\langle k\rangle}(1-2c_{\text{st}})\big[(1-c_{\text{st}})^q+c_{\text{st}}^q\big]}.
    \label{eq:bdepc}
\end{equation}
\end{widetext}
}
In Fig.~\ref{fig:bcdependency}, we present simulated paths in a phase space. 
The starting point is $(c,b)=(1,0)$ for all realizations.
For small networks, as we see, the system fluctuates around stationary values of both concentrations.
However, when the size rises, deviations from Eq.~(\ref{eq:bdepc}) decrease.
For $N=5\cdot 10^5$, phase trajectory lies almost exactly on the theoretical line, and the system ends its evolution in a stationary point.
\blue{Moreover, from Eq.~(\ref{eq:pEq}) we get $p=0$ when $q=1$, and Eq.~(\ref{eq:bdepc}) converges to already derived formula obtained in \cite{vazquez2008analytical} for the linear voter model:
\begin{equation}
    b_{\text{st}}=2c_{\text{st}}(1-c_{\text{st}})\frac{\langle k\rangle-2}{\langle k\rangle-1}.
\end{equation}
}

\section{Phase transitions}
Stationary values of the concentrations depend not only on the model parameters: a group size $q$ and a level of independence $p$ but also on the average node degree of an underlying network $\langle k\rangle$.
In order to better comprehend their impact on the final state of the system, we plot phase diagrams that present these dependencies.
Although we are not able to obtain the direct formula for $c\big(p,q,\langle k\rangle\big)$, we can plot it easily by inverting the figure of $p\big(c,q,\langle k\rangle\big)$, which we obtain from Eqs.~(\ref{eq:pEq}) and (\ref{eq:bdepc}).
The left panel of Fig.~\ref{fig:contPhase} presents phase diagrams of continuous phase transitions, which take place on a random regular graph and Barab\'{a}si-Albert network.
The pair approximation almost exactly predicts the lines of equilibrium and provides better estimation than standard mean-field approach.
\blue{Note that the results for networks with the same average degree match each other despite the fact that investigated structures have very different arrangement of edges.
In the former, all nodes have the same number of neighbors. The latter belongs to the wider class of so called scale-free networks which are characterized by a power law tail in the degree distribution, namely $P(k)\sim k^{-\lambda}$.
In the case of BA model $\lambda=3$ but the analysis can be expanded to other values of the exponent by using alternative network simulation schemes \cite{goh2001universal,fortunato2006scale}.
Herein, we adopt the age ranking method \cite{fortunato2006scale} since it allows the minimal node degree to be adjusted.
The right panel of Fig.~\ref{fig:contPhase} reveals agreement between analytical predictions and simulations for scale-free networks with other tail exponents, as well.

\begin{figure}[!b]
	\centerline{\epsfig{file=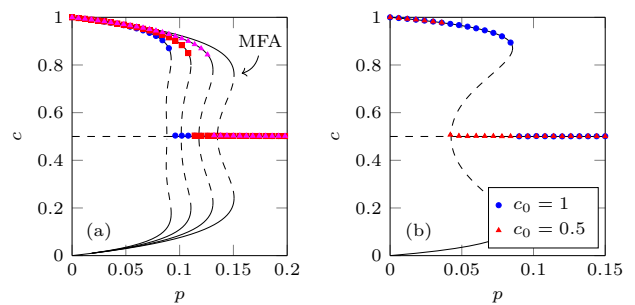,width=1.0\columnwidth}}
	\caption{\label{fig:distPhaseTr}(Color online) The phase diagrams for (a) random regular graph of the size $N=10^5$ and (b) Erd\H{o}s-R\'{e}nyi network with $N=5\cdot10^5$ nodes and $\langle k\rangle=40$. The group of influence consists of $q=6$ agents for RR and $q=8$ for ER. Lines indicate analytical prediction of PA and MFA, solid ones correspond to stable states, and dashed ones refer to unstable points. Markers represent outcomes of Monte Carlo simulations. For the left panel: $\bullet$ $\langle k\rangle=14$, $\scriptstyle\blacksquare$ $\langle k\rangle=20$, $\blacktriangle$ $\langle k\rangle=40$.
		The hysteresis indicates a discontinuous phase transition and matches analytical result.
	}
\end{figure}
Long-range correlations and collective behavior over large scales near criticality are inherent properties of continuous phase transitions \cite{hinrichsen2000non}. Using finite-size scaling (FSS) method \cite{hong2007finite}, presented in the top panel of Fig.~\ref{fig:criticalBeh}, we can estimate the critical level of independence $p^*$. 
Note that we use there standard magnetization $m$ instead of $c$ since it is a proper order parameter; mathematically, it is just re-scaled concentration $m=2c-1$.
Moreover, we can calculate directly the critical point from the obtained pair approximation and compare both results.
}
\blue{At criticality, a system changes its state from ordered $m\neq0$ ($c\neq1/2$) to disordered one $m=0$ ($c=1/2$).}
Therefore, it is enough to investigate the limiting behavior of $p\big(c,q,\langle k\rangle\big)$ when the up spins concentration approaches $1/2$, so calculate the following limit:
\begin{equation}
p^*=\lim_{c\to 1/2} p\big(c,q,\langle k\rangle\big).
\end{equation}
Using previously derived Eqs.~(\ref{eq:pEq}) and (\ref{eq:bdepc}) together with L'H\^{o}pital's rule, we obtain
\begin{equation}
p^*=\frac{q-1}{q-1+2^{q-1}\left(\frac{\langle k\rangle-1}{\langle k\rangle-2}\right)^q}.
\label{eq:criticalPoint}
\end{equation}
\blue{For the system from Fig.~\ref{fig:criticalBeh}, where $q=2$ and $\langle k\rangle=10$, the above formula gives $p^*\approx0.2832$, whereas the FSS technique leads to a bit higher value $p^*=0.2855$.
Nevertheless, the relative difference does not exceed $1\%$; therefore, PA provides better approximation than standard MFA with the critical value $p^*=1/3$, so with the relative error around $17\%$.
In general, the prediction of pair approximation is more accurate when there is a bigger difference between number of individuals in the panel and the average degree of the network.}
When $q$ gets closer to $\langle k\rangle$, some discrepancies arise.
From Eq.~(\ref{eq:criticalPoint}), we also see that the critical level of independence rises together with the growing average degree of a network.
For large $\langle k\rangle$, we can approximate its value by
\begin{equation}
p^*\stackrel{\langle k\rangle\rightarrow\infty}{=} \frac{q-1}{q-1+2^{q-1}},
\end{equation}
which is the result for a complete graph obtained in Ref.~\cite{nyczka2012} by the standard MFA.
\begin{figure}[!t]
	\centerline{\epsfig{file=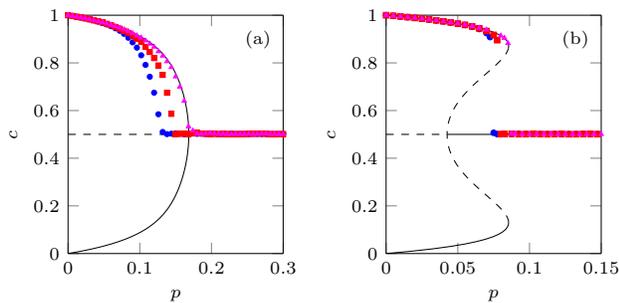,width=1.0\columnwidth}}
	\caption{\label{fig:diagWS}(Color online) Phase diagrams for WS networks of the size $N=2\cdot 10^5$ and with the average degree (a) $\langle k\rangle=8$ and (b) $\langle k\rangle=40$. In the case of a continuous phase transition $q=4$ and for discontinuous one $q=8$.
    With growing rewriting probability $p_r$, the average clustering coefficient $\langle C\rangle$ diminishes, and simulation results approach PA. Around $p_r\approx0.6$ they match each other well. Lines indicate analytical predictions. Markers represent results of Monte Carlo simulations: $\bullet$ $p_r=0.1$, $\scriptstyle\blacksquare$ $p_r=0.2$, $\blacktriangle$ $p_r=0.7$.}
\end{figure}

\blue{The bottom panels of Fig.~\ref{fig:criticalBeh} illustrate the critical behavior of the model. We measure the magnetization and fluctuations $\chi'=N(\Delta m)^2$ as functions of the system size at the critical point.
Since $m(p^*)\sim N^{-\beta/\bar{\nu}}$ and $\chi'(p^*)\sim N^{\gamma'/\bar{\nu}}$, we can estimate the critical exponent ratios \cite{hong2007finite}.
Indeed, for our previously estimated $p^*$, we obtain power-law behaviors of the above quantities in $N$ that confirm the criticality. 
The $q$-voter model acts differently on scale-free networks than Ising model, which, for instance, exhibits an infinite-order phase transition for $2<\lambda\leq3$ \cite{dorogovtsev2008critical}. 
Therefore, it manifests different universal critical behavior.
The analysis can be extended to other numbers of the panel members and SF networks.
}

The pair approximation also works properly in the case of discontinuous phase transitions, see Fig.~\ref{fig:distPhaseTr}.
This time we take into consideration Erd\H{o}s-R\'{e}nyi model, too, as an underlying framework for $q$-voter dynamics.
In the right panel of Fig.~\ref{fig:distPhaseTr}, we can see that the final state depends on the initial conditions. 
This reliance is called hysteresis, and it is one of the characteristic features of discontinuous phase transitions.
\blue{Note that now Eq.~(\ref{eq:criticalPoint}) corresponds to the point $p_1$.
The other point demarcating the upper limit of a region with metastability $p_2$ can be derived numerically by identifying maxima of $p\big(c,q,\langle k\rangle\big)$.}

\begin{figure}[!t]
	\centerline{\epsfig{file=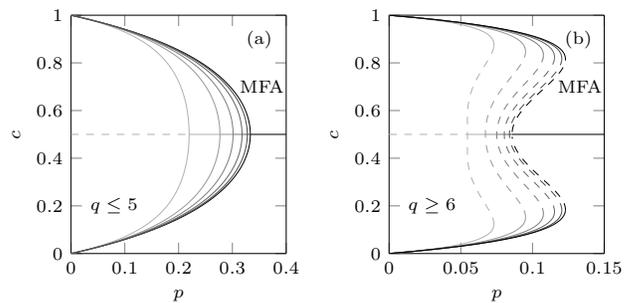,width=1.0\columnwidth}}
	\caption{\label{fig:generalPhase}A schematic representation of the pair approximation for the $q$-voter dynamics with stochastic noise. Darker lines correspond to the higher average degree of a network. The last, black line is an outcome of MFA. The model exhibits two types of phase transitions: (a) continuous and (b) discontinuous.}
\end{figure}

All examined structures above have in common very low clustering coefficient $\langle C\rangle$.
The impact of $\langle C\rangle$ on the model behavior is especially evident for Watts-Strogatz networks, where we can adjust its value by changing the rewriting probability $p_r$.
Note that altering $p_r$ influences only the average clustering coefficient, so the average degree remains constant.
Therefore, PA gives one solution for all rewriting probabilities, which is at odds with simulations; see Fig.~\ref{fig:diagWS}.
However, when $p_r$ is sufficiently large, approximately around 0.6, the clustering coefficient is low enough to neglect dynamical correlations, and above this point all results agree with PA and do not change any more.

In general, PA predicts that given model parameters, only the average degree has an impact on the behavior of a system.
Along with increasing $\langle k\rangle$, phase diagram approaches the one from MFA, see Fig.~\ref{fig:generalPhase}.

\section{Conclusions}
We aimed to investigate the influence of complex networks on the $q$-voter dynamics with stochastic noise.
We put particular emphasis on the analytical approach.
The above model was extensively studied on several complex structures, including Erd\H{o}s-R\'{e}nyi, Barab\'{a}si-Albert, Watts-Strogatz, and scale-free networks.
We also considered its behavior on random regular and complete graphs as a benchmark.
Analytical calculations were supported with a series of Monte Carlo simulations, which allowed us to validate our results.

For the first time, we proposed the pair approximation for the $q$-voter model with independence and confronted it with the broadly used mean-field approach.
In all examined cases we demonstrated the superiority of the former method over the other.
Moreover, our study indicates that for poorly clustered networks pair approximation may provide very accurate description.
We investigated the time evolution of the model and derived an analytical formula for the stationary points, and we analyzed the influence of the system size on fluctuations.
Two types of phase transitions were identified, and we determined the critical level of independence that separates ordered and disordered states \blue{analytically as well as empirically by using finite-size scaling technique.}

On the basis of our analytical and experimental results, we revealed and highlighted the crucial role of the average node degree in predicting the model behavior.
Along with increasing value of $\langle k\rangle$, the phase diagram shifts toward higher values of independence. 
Consequently, it results in higher critical point.
Moreover, for large average degrees we showed that solutions established from PA converge to those from MFA.
In the further studies, in order to improve the prediction accuracy for networks with the average degree close to $q$, one could consider applying the heterogeneous pair approximation \cite{pugliese2009heterogeneous}, which also takes into account degree correlations.

\begin{acknowledgments}
The work was partially supported by funds from the National Science Centre (NCN, Poland) through Grant No. 2013/11/B/HS4/01061.
The author thanks Katarzyna Sznajd-Weron for scientific guidance, meaningful discussions, and suggestions.
\end{acknowledgments}

\end{document}